# Secondary Use of Health Data: Centralized Structure and Information Security Frameworks in Finland

*Hannu Vilpponen*[1,2], *Antti Piirainen*[2], *Miikka Kallberg*[3], *and Tommi Mikkonen*[1]

[1] University of Jyväskylä, Finland
[2] Finnish Social and Health Data Permit Authority Findata, Finland
[3] CSC – IT Center for Science Ltd., Finland
```
{hannu.v.vilpponen, tommi.j.mikkonen}@jyu.fi,
antti.piirainen@findata.fi, miikka.kallberg@csc.fi
```

**Abstract.** The utilization of health data for secondary purposes, such as research, statistics, and development, has become increasingly significant in advancing healthcare systems. To foster the above, Finland has established a framework for the secondary use of health data through legislative measures and the creation of specialized institutions, which are the first of their kind in the world. In this paper, we give an overview of our implementation for using secondary health data in a centralized fashion. As a technical contribution, we also address key implementation aspects related to implementing the framework.

## 1 Introduction

The utilization of health data for secondary purposes, such as research, statistics, and development, has become increasingly significant in advancing healthcare systems. In April 2024 the European Parliament adopted an agreement for European Health Data Space (EHDS) regulation [6]. The aim of the EHDS is to make it easier to access and exchange health data across borders (primary use of data) and inform health research and policy-making (secondary use of data) [5]. The benefits of EHDS include empowering individuals to manage their health data, supporting the use of health data to improve healthcare services, research, innovation, and decision-making, and providing the EU with the means to safely harness the potential for the seamless exchange, use and re-use of health data. Finland, known for its robust healthcare infrastructure and innovative data practices, has established a framework for the secondary use of health data through legislative measures and the creation of specialized institutions.

The Nordic countries have a long tradition of population-based health registers. Record-keeping in the Nordic countries started in the 1950s [7]. While the Nordic registries were established at various times, they generally follow a similar overall structure. Each country has a significant number of registries and databases, with Denmark having over 200 alone [7] and Finland having several hundred as well. In Finland, the recognition of the need for more streamlined, secured and standardized secondary use of health data led to the establishment of the necessary authority, Findata, in 2019. Findata is the data permit authority in the social welfare and healthcare sector, and its assigned role is to organize and maintain a centralized contact point for health data for researchers [2].

In this paper, we outline the framework for the secondary use of health data in Finland. In doing so, we address the role of Findata in defining and implementing data usage environments. As a technical contribution, we also consider key implementation aspects. The rest of this paper is structured as follows. In Section 2, we present the background and motivation for the work. In Section 3, we outline the framework used for information security in Finland. In Section 4, we discuss the technical implementation. Finally, in Section 5, we draw some final conclusions.

## 2    Background and Motivation

The secondary use of health data has been studied in several reports [8, 9, 10, 11, 32]. For example, "Operationalizing Research Access in Platform Governance" by Algorithm Watch [8] identifies best practices for research access in platform governance by learning from legal frameworks in environmental law and medical research. It highlights the need for better data access to ensure accountability, using two case studies to address transparency challenges and data protection concerns, from which the other showcases an in-depth examination of Findata's operations.

Center for Data Innovation's report "How the EU Can Unlock the Private Sector's Human-Mobility Data for Social Good" by Hodan Omaar [9] has a different goal, as it discusses how businesses collect mobility data, and how this data can be valuable for addressing societal challenges such as disease spread, urban planning, and disaster response. While researchers and governments benefit from such data, access is often limited due to privacy concerns, high costs, and legal uncertainties. As an example of national legislation which takes on to tackle these issues, the report examines Findata and the Finnish Act on Secondary Use of Health and Social Data.

Nuffield Trust's "Fit for the future: What can the NHS learn about digital health care from other European countries?" [10] examines how digital health care is crucial for sustainable healthcare systems and improving public health. By studying five European countries — Denmark, Finland, Sweden, Estonia, and Portugal — the report highlights that countries with strong digital policies and public trust in digital services have made significant progress. Key factors include effective governance, public confidence, and collaborative design of digital tools.

The Sanitas Health Insurance Foundation's study, "Are smartwatches eroding solidarity?" [11] explores future healthcare scenarios, reflecting on the impact of data-driven health systems on societal cohesion and solidarity. It uses Finland as an example of government-driven action regarding data utilization.

VTT Technical Research Centre of Finland has studied the subject from a more technical point of view and concentrates on secure processing environments. The report [32], commissioned by the Finnish Ministry of Social Affairs and Health, examines the Act's impact on AI research through interviews with stakeholders, exploring challenges in developing secure processing environments. It identifies technological solutions and provides recommendations for efficient data use in research.

Despite the comprehensive studies, such as the aforementioned and others in the field, the structure and information security frameworks and key implementation aspects have not yet been thoroughly addressed.

Furthermore, the European Health Data Space (EHDS) initiative is expected to enhance cross-border data sharing and collaboration, amplifying the potential for secondary use of health data at a European level. This will pave the way for more comprehensive research opportunities and healthcare improvements across Europe, fostering a unified approach to health data utilization and innovation. Presently, there are several recent projects that aim to define security and design principles for research environments for health data, listed in Table 1.

Table 1. Projects to define research environments for health data in Europe.

| | |
|---|---|
| EOSC ENTRUST [24] | EOSC-ENTRUST aims to create a European network of Trusted Research Environments (TREs) for sensitive data and drive European interoperability by joint development of a common blueprint for federated data access and analysis. |
| TEHDAS & TEHDAS2 [25] | The TEHDAS Joint Action provided background information for the preparation of the EHDS regulation proposal. TEHDAS2 Joint Action is doing the same for the implementing acts of the EHDS and prepares the ground for the harmonized implementation of the secondary use of health data. |
| TRE (UK) [26] | TRE (UK) aims to prepare a blueprint for Trusted Research Environment in the UK for researching Health data. |
| Healthdata@EU Pilot [28] | The HealthData@EU Pilot project brought together 17 partners including health data access bodies, health data sharing infrastructures and European agencies. It will build a pilot version of the EHDS infrastructure for the secondary use of health data which will serve research, innovation, policy making and regulatory purposes. |
| EHDS2 Communities of Practice [27] | The HDABs-CoP mission is to foster collaboration and knowledge sharing among Competent Authorities and Affiliated Entities involved in establishing the HDABs and responsible for the secondary use of health data within the EHDS. |

## 3 Framework for Secondary Use of Health Data in Finland

The Act on the Secondary Use of Health and Social Data provides the legal foundation for the secondary use of health data in Finland [1]. The legislation aims to facilitate the use of social welfare and healthcare data while ensuring data protection and privacy. It defines the conditions under which health data can be used for purposes beyond direct patient care, such as scientific research, statistics, and policy development. Findata has established a regulation on Secure Processing Environments (SPE), which is used as one of the audit criteria in the audit processes of SPE's [3]. The primary objectives of the framework for Secondary Use of Health Data include:
- Enhancing the quality and effectiveness of health and social services.
- Promoting research and innovation.
- Ensuring the security and privacy of health data.
- Facilitating the efficient and ethical use of health data for secondary purposes.

Findata, the Finnish Social and Health Data Permit Authority [2], was established in 2019 under the Act on the Secondary Use of Health and Social Data and is the national permit authority. Findata's responsibilities include:
- Granting permits for the secondary use of health and social data.
- Compiling and pre-processing the data stated in permits and data requests.
- Ensuring compliance with data protection regulations.
- Providing secure data environments for data transfer, processing and analysis.

**Data Usage Environments.** Findata grants data permits, collect, links and pseudonymises the data and produces aggregated data on request. It plays a crucial role in defining the environments where health data can be accessed and analyzed. These SPEs are designed to provide secure and controlled access to data, ensuring that data privacy and integrity are maintained. On the other hand, the service providers of SPEs are under constant pressure to meet the demand for up-to-date software and performance. Research data can include numerous variables from several registers, even covering the entire population along with comparison data. The size of a single SPE can reach several terabytes.

According to the legislation, Findata is tasked with issuing specific requirements for SPEs, against which security audits are conducted to ensure compliance. Findata must provide a SPE of its own, but other organizations, both public and private, can build audited SPEs as well. In this way, they can manage SPE system configuration and costs independently [12,13,14,15,16,17,18,19,20,21]. In Finland, there are currently ten audited SPEs, whose compliance with the law is overseen by Valvira, the National Supervisory Authority for Welfare and Health [4]. SPEs established today are listed in Table 2.

Table 2. Secure Processing Environments and their approximate volumes in Finland.

| Name of the SPE | Provider | Ownership | Active environments | Active users |
|---|---|---|---|---|
| HUS Acamedic | HUS Helsinki University Hospital | Wellbeing services county | 406 | 1 300 |
| T3 Researchers workspace | Istekki Oy | Publicly-owned private company | 112 | 354 |
| SPESiOR | ESiOR Oy | Private company | n/a | n/a |
| FIMM Sandbox | University of Helsinki | University (public) | 1 | 2 |
| FinnGen Sand-Box | University of Helsinki | University (public) | 30 | 816 |
| Kapseli | Findata | Public authority / Government | 145 | 1 000 |
| FIONA remote access system | Statistics Finland | Public authority / Government | 301 | 1 204 |
| SD Desktop | CSC - IT Center for Science Ltd | Publicly-owned special interest company | 21 | 65 |
| SECDATA | Aalto University | University (public) | 4 | 10 |
| Auria's Atolli | Auria Biobank | Wellbeing services county | 55 | 260 |
| Total | | | 1 075 | 5016 |

## 4    Technical Implementation

**Secure Data Processing Environments.** The technical implementation of secure data environments involves the development of platforms and tools that allow researchers and other authorized users to access and analyze data without compromising security. Key features of these environments include:
- **Data Anonymization and Pseudonymization:** Before transferring health data to SPEs, permit authorities use safeguard measures to protect personal identities while allowing data analysis.
- **Isolation**: The SPE is isolated from other parts of the system, such as the operating system or applications running on the main processor. This isolation helps prevent unauthorized access or tampering with the sensitive data processing environment.
- **Access Control and Monitoring:** Systems log control and monitor who accesses the data and how it is used.

- **Encryption:** Ensures that data is encrypted both in transit and at rest to prevent unauthorized access.

**Findata Kapseli.** CSC – IT Center for Science Ltd., plays a pivotal role in the technical implementation by providing a secure data environment called Kapseli. Kapseli is designed to meet stringent data protection requirements and supports the secure processing of sensitive health data.

**Kapseli Security.** Kapseli is audited by an external auditor to match Findata's requirements for a secure data environment [3]. In Finland, there are two companies that are authorized by the National Cyber Security Center to audit the regulation on the requirements set for SPEs. The regulation is based on the existing National Security Auditing Criteria (KATAKRI) framework and legislation of secondary use of health data. Kapseli has been built using existing security recommendations including Microsoft Admin Tier model [22], MFA and other well-known security frameworks, such as CIS Benchmarks [23]. The CIS Benchmarks™ are prescriptive configuration recommendations for more than 25+ vendor product families. They stand for the consensus-based effort of cybersecurity experts globally to help you protect your systems against threats more confidently.

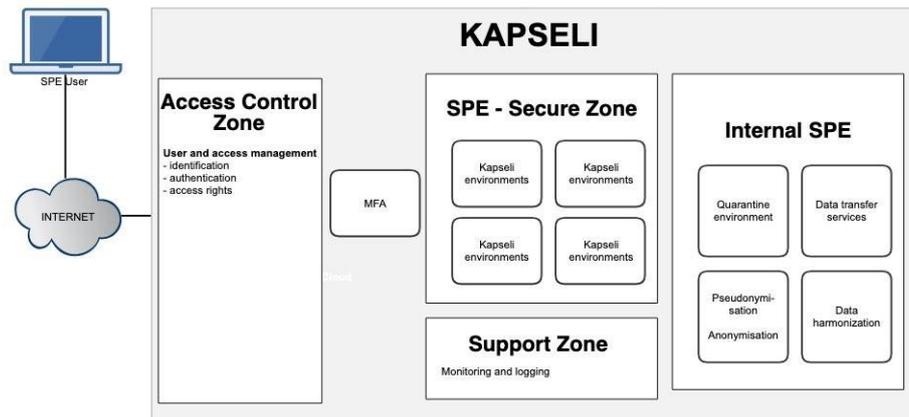

**Fig. 1.** Kapseli high-level architecture.

**Kapseli Architecture**. The architecture of Kapseli (Fig. 1) ensures robust security and efficient data management. In terms of the underlying implementation, Findata's Kapseli environment is built on top of CSC ePouta -private cloud. ePouta is designed for processing sensitive data and is located in Finland. Key Kapseli components include the following:

- **Data Isolation:** Each research project operates in a separate, isolated environment to prevent data leakage and unauthorized access.
- **High-Performance Computing Resources:** Kapseli leverages CSC's high-performance computing infrastructure to support large-scale data analysis.
- **Scalable Storage Solutions:** The environment is equipped with scalable storage options to accommodate varying data sizes and types.

Kapseli is partitioned to a number of different zones, called Access Control Zone, SPE-Secure Zone, Support Zone and Internal SPE. The Access Control Zone contains services for user authentication and identification. Access to this zone is limited to requested IP addresses. Users are identified by using existing identification federations, for example Suomi.fi [30], which utilizes banks' electronic identification for user identification in an eIDAS [29] compliant way. Other supported federations are Haka, the identity federation of the Finnish universities, polytechnics and research institutions [31], and Virtu, the Finnish Government's joint single sign-on solution, provided by the Government ICT Centre Valtori [32]. After successful identification, users are redirected to Multifactor authentication before accessing the Kapseli environment in the SPE-Secure zone.

The SPE-Secure Zone contains all Kapseli environments. Currently there are approximately 145 permit specific user environments in the SPE-Secure Zone of Kapseli. Each Kapseli environment is isolated from other Kapseli environments and from the internet. Kapseli environments can be either Windows or Ubuntu Linux virtual machines. Findata maintains a list of pre-approved software that can be installed on Kapseli environments, some are installed by default and others as requested. Users don't have administrative permissions to Kapseli environments. All data and software must pass through Findata's inspection before it can be brought to Kapseli environments. The research data is provided to specific Kapseli environments by Findata.

The Support Zone contains necessary support functions to ensure data security and privacy, such as monitoring services on other zones and logging. Internal SPE zone is dedicated to the use of Findata officials for processing the health data before it can be provided to the Kapseli environment, like pseudonymisation, anonymisation and data harmonization. It also contains necessary services for transferring the data and tools to check data anonymization before it is released outside the Kapseli environment.

## 5 Discussion and Conclusions

The secondary use of health data in Finland is supported by a robust legislative framework and the dedicated efforts of Findata and CSC. Through secure data environments such as Kapseli and comprehensive support systems, Finland ensures that health data can be used effectively for research and development while maintaining the highest standards of data protection and privacy. The successful implementation of these systems highlights Finland's commitment to leveraging health data for the betterment of healthcare services and public health.